\documentclass[12pt,epsf]{article}
\usepackage{graphicx}
\usepackage{amsmath}
\usepackage{amssymb}
\usepackage{color}

\usepackage{multirow}
\usepackage{rotating}
\usepackage{subfig}

\newcommand{\be}{\begin{equation}}
\newcommand{\ee}{\end{equation}}
\newcommand{\bea}{\begin{eqnarray}}
\newcommand{\eea}{\end{eqnarray}}
\newcommand{\beas}{\begin{eqnarray*}}
\newcommand{\eeas}{\end{eqnarray*}}
\newcommand{\ba}{\begin{array}}
\newcommand{\ea}{\end{array}}

\newcommand{\nbox}{{\,\lower0.9pt\vbox{\hrule \hbox{\vrule height 0.2 cm \hskip 0.19 cm \vrule height 0.2 cm}\hrule}\,}}

\def\href#1#2{#2}

\begin{document}
\begin{titlepage}
\hfill

\vspace*{20mm}
\begin{center}
{\Large \bf Acceleration-Induced Deconfinement Transitions \\ in de Sitter Spacetime}

\vspace*{15mm}
\vspace*{1mm}
Jonathan Blackman, Michael B. McDermott, Mark Van Raamsdonk\footnote{e-mails: blackman@interchange.ubc.ca, mbmcderm@phas.ubc.ca, mav@phas.ubc.ca}
\vspace*{1cm}

{\it Department of Physics and Astronomy,
University of British Columbia\\
6224 Agricultural Road,
Vancouver, B.C., V6T 1W9, Canada\\
}

\vspace*{1cm}
\end{center}

\begin{abstract}
In this note, we consider confining gauge theories in $D=2,3,4$ defined by  $S^2$ or $T^2$ compactification of higher-dimensional conformal field theories with gravity duals. We investigate the behavior of these theories on de Sitter spacetime as a function of the Hubble parameter. We find that in each case, the de Sitter vacuum state of the field theory (defined by Euclidian continuation from a sphere) undergoes a deconfinement transition as the Hubble parameter is increased past a critical value. In each case, the corresponding critical de Sitter temperature is smaller than the corresponding Minkowski-space deconfinement temperature by a factor nearly equal to the dimension of the de Sitter spacetime. The behavior is qualitatively and quantitatively similar to that for confining theories defined by $S^1$ compactification of CFTs, studied recently in arXiv:1007.3996.

\end{abstract}
\vskip 2cm

\begin{center}
\vskip 1cm

\end{center}
\end{titlepage}

\vskip 1cm
\vskip 0.1 in
\noindent

\section{Introduction}

\subsubsection*{Motivation and Background}

The study of quantum field theory on curved spacetime is an important subject, crucial for our understanding of black hole physics and early universe cosmology. Most of the well-known results, from the derivation of Hawking radiation to the study of quantum fluctuations of fields during inflation, are obtained simply by studying free fields (see, for example \cite{Birrell:1982ix}). However, since realistic quantum field theories are interacting, it is interesting to investigate new phenomena that may arise for interacting quantum field theories on nontrivial spacetimes. This presents new technical challenges even for weakly-coupled quantum field theories, while for strongly coupled field theories on curved spacetime, either direct analytic analysis or numerical simulation is not currently possible.

Fortunately, a powerful alternative method for studying certain strongly-coupled field theories is now available in the AdS/CFT correspondence \cite{Maldacena1998}. In the form that we will use in this paper, this suggests that certain conformal field theories on some fixed spacetime manifold $M$ are equivalent to gravitational theories on asymptotically locally AdS spacetime with boundary geometry $M$. In most studies, the geometry $M$ has been chosen to be Minkowski space or $S^d \times R$, but we are free to take $M$ to be a cosmologically interesting background such as de Sitter space.

In recent work \cite{Marolf:2010tg} (see also \cite{Aharony:2002cx, Balasubramanian:2002am, Cai:2002mr, Ross:2004cb, Balasubramanian:2005bg, He:2007ji, Hutasoit:2009xy}), the AdS/CFT correspondence was used to show that certain confining gauge theories on de Sitter spacetime undergo a deconfinement transition at a critical value of the acceleration parameter (i.e. the Hubble constant) for the de Sitter space. To study confining rather than conformal field theory, the authors used the observation \cite{Witten:1998zw} that a conformal field theory compactified on a circle with antiperiodic boundary conditions for fermions defines a confining field theory in one lower dimension with particles and a mass gap. On Minkowski space, such theories undergo a deconfinement transition at some critical temperature $T_c$. In de Sitter space, it is well known \cite{Gibbons:1977mu} that there is an effective temperature $T_{dS}$ related to the Hubble constant by\footnote{This is the temperature registered by a thermometer interacting with free fields in their de Sitter-invariant vacuum state on de Sitter space.}
\[
T_{dS} = {H \over 2 \pi} \; .
\]
The thermal character of physics on de Sitter space suggests that for large enough $H$, confining gauge theories should exist in a deconfined plasma phase rather than the hadronic phase that exists at low temperatures. The result of \cite{Marolf:2010tg} is that for the confining theories on $dS^d$ defined by CFTs on $dS^d \times S^1$, this deconfined phase is favored when
\be
\label{res1}
T_{dS} > {1 \over d} T_c \; .
\ee
Thus, deconfinement in de Sitter spacetime occurs for a de Sitter temperature that is lower than the the Minkowski spacetime transition temperature by a factor of the spacetime dimension.

\subsubsection*{Summary}

The main goal of this paper is to study more examples of confining gauge theories on de Sitter space, to investigate the phase structure as a function of the field theory parameters and acceleration parameter and to determine whether (\ref{res1}) represents a general result. We consider two new examples, conformal field theory compactified on $T^2$ and conformal field theory compactified on $S^2$. In each case (assuming that the boundary conditions for fermions are antiperiodic on at least one cycle of the $T^2$), we obtain a confining gauge theory in two lower dimensions.\footnote{Our results apply to any conformal field theory whose gravity dual has a consistent truncation to pure Einstein gravity with negative cosmological constant. As a specific example of a theory that will lead to a four-dimensional confining gauge theory, we can consider the $(0,2)$ conformal field theory in six dimensions that describes the low-energy physics of M5-branes.}

Physical properties of quantum field theory in the canonical de Sitter vacuum state are related by analytic continuation to the physics of the corresponding Euclidean theory on a sphere,\footnote{For a recent review, see \cite{Spradlin:2001pw}.} where the radius of the sphere is related to the acceleration parameter of the de Sitter space by $R_{S^d} = 1/H$. Thus, our problem is equivalent to investigating the phase structure of a conformal field theory on $S^2 \times S^d$ or $T^2 \times S^d$. In the first case, we have only a single parameter $R_{S^2}/R_{S^d}$ (the acceleration parameter in units of the inverse $S^2$ radius), while in the second case, we have two parameters $R_1/R_{S^d}$ and $R_2/R_{S^d}$ corresponding to the sizes of the two cycles of the torus (which we take to be rectangular). For each choice of parameters, the phase of the field theory with lowest free energy corresponds to the gravitational solution with minimum action whose boundary metric matches with the field theory geometry.

After choosing a metric ansatz with the appropriate symmetries and imposing the boundary conditions, we find in each case that the possible solutions are determined by a system of coupled ordinary differential equations than can be solved numerically. In some cases, multiple solutions exist for a given boundary geometry, so we must numerically evaluate the gravitational action to find which one is minimal. In each solution, either the $S^d$ or the compactification manifold ($T^2$ or $S^2$) degenerates at some point along the radial direction away from the boundary. As discussed in detail in \cite{Marolf:2010tg}, geometries for which the $S^d$ shrinks to zero correspond to deconfined physics in the dual field theory, while geometries for which the $T^2$ or $S^2$ shrinks to zero correspond to confined physics. The boundary between these two behaviors in the phase diagram represents an acceleration-induced deconfinement transition.

\begin{figure}[t]
  \begin{center}
 \scalebox{0.4}{ \includegraphics{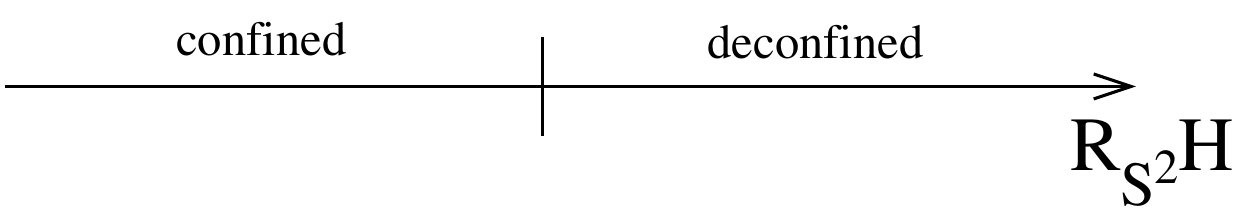}}
  \end{center}
   \caption{De Sitter space phase diagram for confining gauge theory defined by $CFT_{d+2}$ on $S^2$.}
   \label{pd2}
\end{figure}

Our results for the phase diagrams are summarized in figures \ref{phase_diagram} and \ref{pd2}. In each case, we find that the theory deconfines for sufficiently large $H$. In table \ref{table:summary}, we compare the de Sitter temperature at the transition with the Minkowski-space deconfinement temperature. We find that the critical de Sitter temperature is close to but not exactly given by the result (\ref{res1}).

\begin{table}[t]
\centering
\vskip 0.1 in
\begin{tabular}{| c |  c | c | c | c| }
\hline
Theory & Dimension & Minkowski $T_c$ & critical $T_{dS}$ & $T_c/T_{dS}$ \\   \hline   \, & &&&\\

$CFT_4$ on $S^2$ & 2 & $0.336 R_{S^2}^{-1}$ & ${1 \over 2 \pi} R_{S^2}^{-1}$ & 2.11\\[0.1in]
$CFT_4$ on $T^2$ & 2 & ${1 \over 2 \pi} R_{min}^{-1}$ & ${0.474 \over 2 \pi} R_{min}^{-1}$ & 2.11\\[0.1in]
$CFT_3$ on $S^1$ & 2 & ${1 \over 2 \pi} R_{S^1}^{-1}$ & ${1 \over 4 \pi} R_{S^1}^{-1}$ & 2\\[0.1in]
\hline &&&& \\
$CFT_5$ on $S^2$ & 3 & $0.351 R_{S^2}^{-1}$  & $ 0.115 R_{S^2}^{-1}$ & 3.05 \\[0.1in]
$CFT_5$ on $T^2$ & 3 & ${1 \over 2 \pi} R_{min}^{-1}$ & ${0.317 \over 2 \pi} R_{min}^{-1}$& 3.15 \\[0.1in]
$CFT_4$ on $S^1$ & 3 & ${1 \over 2 \pi} R_{S^1}^{-1}$ & ${1 \over 6 \pi} R_{S^1}^{-1}$ & 3\\[0.1in]
\hline &&&& \\
$CFT_6$ on $S^2$ & 4 & $0.365 R_{S^2}^{-1}$  & $ 0.093 R_{S^2}^{-1}$ & 3.92 \\[0.1in]
$CFT_6$ on $T^2$ & 4 & ${1 \over 2 \pi} R_{min}^{-1}$ & ${0.239 \over 2 \pi} R_{min}^{-1}$& 4.18 \\[0.1in]
$CFT_5$ on $S^1$ & 4 & ${1 \over 2 \pi} R_{S^1}^{-1}$ & ${1 \over 8 \pi} R_{S^1}^{-1}$ & 4\\[0.1in]

\hline
\end{tabular}
\caption{Minkowski and de Sitter space deconfinement temperatures for confining gauge theories defined by compactified CFTs}
\label{table:summary}
\end{table}

\section{Generalities}

To begin, we briefly review the AdS/CFT approach to studying field theories on non-trivial spacetime metrics (more discussion can be found in \cite{Marolf:2010tg, Hubeny:2009ru,Hubeny:2009kz,Hubeny:2009rc}). In this paper, we consider conformal field theories dual to a gravitational theory that has Einstein gravity with negative cosmological constant as a consistent truncation. Examples include the four-dimensional ${\cal N}=4$ SYM theory arising from D3-branes in string theory and the six-dimensional $(0,2)$ CFT arising from M-theory fivebranes. We define lower dimensional confining theories by compactification on some manifold $M$. We are interested in studying the equilibrium physics of these lower-dimensional theories either on Minkowski space with a finite temperature, or on de Sitter space in the Euclidean vacuum state. The physics of these theories can be obtained by studying the Euclidean field theory on $R^{d-1} \times S^1$ and $S^{d}$ respectively. Thus, in the end we are interested in properties of a Euclidean CFT on a fixed space ${\cal B}$ of the form $M \times S^1 \times R^{d-1}$ or $M \times S^{d}$. In particular, we would like to investigate possible phase transitions that occur as the geometrical parameters of these spaces are varied.

According to the AdS/CFT correspondence, the Euclidean field theory on a space ${\cal B}$ is dual to the minimum-action solution of the gravity theory which is asymptotically locally AdS and whose boundary metric is the metric of ${\cal B}$.\footnote{In general, the path integral on the field theory side maps to the path integral on the gravity side, but we are considering a limit for which the gravity theory is well-approximated by the classical low-energy gravity. In this case, the gravitational path integral is dominated by the minimum action solution.} Explicitly, we demand that the spacetime metric may be written as
\be
\label{alads}
ds^2 = l^2({dr^2 \over r^2} + r^2 g_{ij}(r,x) dx^i dx^j)
\ee
where for large $r$,
\be
\label{asympt}
g_{ij}(r,x) = g^{\cal B}_{ij}(x) + {\cal O}({1 \over r}) \; .
\ee
We will always assume that the preferred field theory state does not break any of the geometrical symmetries of the background ${\cal B}$. In this case, the corresponding bulk metric must also preserve these symmetries, and we can consider an ansatz for $g_{ij}(r,x)$ that reflects this. In all the cases we consider, the symmetries are enough to specify $g_{ij}(r,x)$ up to a small number of undetermined functions of $r$. These functions must satisfy coupled ordinary differential equations that follow from Einstein's equations
\be
\label{EE}
{\cal R}_{\mu \nu} - {1 \over 2} g_{\mu \nu} {\cal R} - {(D-1)(D-2) \over 2 l^2} g_{\mu \nu} = 0
\ee
where $D$ is the spacetime dimension of the gravity theory and $l$ is the AdS radius. Demanding the asymptotic behavior (\ref{asympt}) provides a set of boundary conditions for the functions at large $r$, while the remaining boundary conditions come from demanding that the Euclidean geometry is smooth in the interior (e.g. avoiding conical singularities). The differential equations are generally too complicated for an analytic solution, but it is straightforward to integrate them numerically.\footnote{In performing numerics, it is necessary to treat carefully the IR boundary of the geometry, where one of the metric functions goes to zero. At these places it can be helpful to determine a power series solutions to the equations near the point $r=r_0$, and use the series solutions to determine the values of the functions and their derivatives at some point $r=r_0 + \epsilon$ where $\epsilon>0$. The differential equations can then integrated numerically from $r=r_0+\epsilon$.}

In some cases, there exist multiple smooth solutions with the same asymptotic behavior. Here, the different solutions represent different possible phases of the field theory. The phase with lowest free-energy corresponds to the gravity solution with smallest action. Thus, we need to compare the gravitational action for the various solutions to determine which is the smallest. In general, this has divergences from integrating over any non-compact directions in the field theory and also from integrating over the radial direction in the bulk. Thus, we should compute the action per unit field theory volume, and introduce a regularization scheme to deal with the radial divergence. The standard procedure is to work with a cutoff surface in the bulk defined at some radial position $r=r_M$.

The gravity action including the boundary term is\footnote{Here, $\gamma$ is the metric induced on the boundary surface $r = r_M$, and $K$ is trace of the second fundamental form of the boundary metric, defined as
\[
K = \gamma^{\mu \nu} \nabla_\mu n_\nu \;,
\]
where $n^\mu$ is the outward unit normal vector at $r=r_M$.}
\be
\label{action}
S = \frac{1}{16 \pi G^{(D)}} \left[ -  \int_{\cal M} d^{D} x \sqrt{g} \left\{ {\cal R} + {(D-1)(D-2) \over \ell^2}  \right\} -2 \int_{\partial \cal M} d^{D-1} x \sqrt{\gamma} K \right] \;,
\ee
but this expression diverges as $r_M \to \infty$. The divergences are generally cured if we introduce a counterterm action defined in terms of the intrinsic geometry of the boundary surface \cite{Balasubramanian:1999re,Emparan:1999pm},
\bea
\label{ctaction}
S_{ct} &=& \frac{1}{16 \pi G^{(D)}} \int_{\partial \cal M} d^d x \sqrt{\gamma} \left\{ 2(D-2) + {1 \over (D-3)} \hat{{\cal R}}  + {1 \over (D-3)^2(D-5)} \hat{\cal R}_{ab} \hat{\cal R}^{ab} \right. \cr
&& \qquad \qquad \qquad \qquad \qquad \qquad \qquad  \left.  - {D-1 \over 4 (D-2) (D-3)^2 (D-5)} \hat{\cal R}^2 \right\} \;,
\eea
where $\hat{\cal R}$ is the intrinsic curvature of the boundary metric and we use only the first term for $D=3$, the first and second terms for $d=4,5$ and all terms for $d=6,7$. The counterterms generally render the action density finite, but for certain geometries in even $d$ there remains a logarithmic divergence related to the presence of a conformal anomaly in the boundary field theory. In these cases, we can either subtract off a logarithmic counterterm or compare the regulated actions for two solutions with the same boundary geometry before taking the cutoff to infinity. In such a comparison, the choice of cutoff surface for the two geometries should be based on the same geometric criterion (e.g. at some fixed radius for a particular circle or sphere). In this case, the action difference approaches a finite limit as the cutoff is taken to infinity. A detailed example of the comparison between actions for various solutions is given in appendix B.

Phase transitions occur at points where the action for one family of solutions becomes smaller than the action for another family of solutions as the geometrical parameters of the boundary metric are varied.

\section{Confining gauge theories from compactified CFTs}

In this section, we introduce the various confining gauge theories that we study in this paper and calculate the Minkowski space deconfinement temperatures. In each case, we begin with a conformal field theory dual to a gravitational theory that has Einstein gravity with negative cosmological constant as a consistent truncation. The confining gauge theory is defined by compactifying this theory on a manifold $M$ that we will choose to be $T^2$ or $S^2$. The equilibrium finite temperature physics on Minkowski space is governed by the behavior of the Euclidean CFT on $M \times S^1 \times R^{d}$, where the $S^1$ represents the Euclidean time direction, compactified with a period $1/T$.

As discussed in the previous section, the phase with lowest free energy corresponds to the minimum-action asymptotically locally AdS gravity solution with boundary metric $M \times S^1 \times R^{d}$. For sufficiently small $S^1$ (i.e. high-enough temperature), the minimum action solution is always one where the $S^1$ contracts smoothly to zero in the interior of the geometry. The corresponding Lorentzian solution is a black brane solution with a horizon, and the corresponding field theory state is deconfined \cite{Witten:1998zw}. For low temperatures (large $S^1$ radius) the minimum action solution is one for which the manifold $M$ contracts to zero size in the interior of the geometry while the $S^1$ size remains finite.\footnote{This assumes that the spin structure on $M$ is such that $M$ can be the boundary of a smooth spin manifold. This is automatic for $S^2$ but requires at least one cycle with antiperiodic boundary conditions for fermions in the $T^2$ case.} This solution ends smoothly in the IR part of the geometry and corresponds to a confined state of the field theory.

\subsection{$CFT_{d+2}$ on $T^2$}

For the theory defined by compactification on $T^2$, the analysis is particularly simple, since there is no distinction between the thermal circle (which has antiperiodic boundary conditions for fermions) and the circles on the torus with antiperiodic boundary. Each of these circles is allowed to pinch off smoothly in the interior. The solution with boundary metric $T^2 \times S^1 \times R^{d-1}$ for which some $S^1$ pinches off is obtained by taking the AdS soliton metric (the minimum action solution with boundary geometry $S^1 \times R^{d+1}$), and periodically identifying two of the $R^{d+1}$ directions. Solutions constructed in this way are dual to deconfined or confined phases depending on whether the contracting circle from the AdS soliton is identified with the thermal circle or with a $T^2$ circle.

The minimum action solution is the one for which the smallest boundary circle contracts in the interior. Thus, we have a deconfinement transition at critical temperature
\[
T_c = {1 \over 2 \pi R_{min}}
\]
i.e. at the temperature for which the thermal circle becomes smaller than the smallest cycle on the torus.

\subsection{$CFT_{d+2}$ on $S^2$}

To determine the Minkowski-space deconfinement temperature for the $d$-dimensional theory defined by a $CFT_{d+2}$ compactified on $S^2$, we need to study asymptotically AdS solutions with boundary geometry $S^2 \times S^1 \times R^{d-1}$. As the $S^1$ radius is decreased relative to the $S^2$ radius, we expect a phase transition between a solution with $S^2$ contractible in the bulk and a solution with contractible $S^1$. For $d=2$, the relevant geometries were analyzed in \cite{Copsey2006}, and we extend that analysis to more general $d$ in appendix A. The results for the deconfinement transition temperature for $d=2,3,4$ are:
\be
\label{S2trans}
T_c = \left\{ \ba{ll} 0.336 R_{S^2}^{-1}  &  \qquad d=2 \cr  0.351 R_{S^2}^{-1}  &  \qquad d=3 \cr 0.365 R_{S^2}^{-1} & \qquad d=4 \ea \right. \; .
\ee

\section{Confining gauge theories on de Sitter spacetime}

We now proceed to study the confining gauge theories defined in the previous section on a de Sitter spacetime background. We are interested in the physics of the theory in the canonical de Sitter-invariant vacuum state. Physical observables in this state of the field theory on $dS^{d}$ may be defined by analytic continuation from the Euclidian theory on the sphere $S^{d}$. Thus, we would like to investigate the behavior of the Euclidean CFT on $M \times S^{d}$ for $M$ equal to $T^2$ or $S^2$. By the AdS/CFT correspondence, the physics can be obtained by determining the minimum-action asymptotically AdS solution with boundary metric $M \times S^{d}$.

As for the finite-temperature Minkowski space case, there are two topologically-distinct types of solutions. Solutions for which the $S^{d}$ pinches off smoothly in the interior of the geometry correspond to Lorentzian solutions with horizons (similar to the topological black hole solution \cite{Banados:1997df}) and are dual to deconfined field theory states. Solutions for which the compactification manifold $M$ is contractible in the bulk while the $S^d$ is not are related to horizon-free Lorentzian solutions (similar to the AdS ``bubble of nothing'' \cite{Aharony:2002cx}) that are dual to confined field theory states.

\subsection{CFT on $T^2 \times S^d$}

For a $(d+2)$-dimensional CFT compactified on $T^2$, the radii of the two boundary circles of the torus will be denoted $R_{\chi}$ and $R_{\xi}$.  Conformal invariance guarantees that the physical properties of this theory on de Sitter space with acceleration parameter $H$ depend only on the dimensionless parameters $R_{\chi} H$ and $R_{\xi} H$, or equivalently $R_{\chi} / R_{S^d}$ and $R_{\xi} / R_{S^d}$.

We will assume that the field theory state does not spontaneously break any of the geometrical symmetries. Thus, the field theory state and the corresponding gravity solution should possess an $SO(d+1) \times U(1)\times U(1)$ symmetry.

The most general dual metric with $SO(5) \times U(1)\times U(1)$ symmetry can be put in the form:
\begin{equation}
\label{genpar}
ds^2= A(\rho) d\rho^2   + B(\rho) d\chi^2 + C(\rho) d\xi^2 + D(\rho) d\Omega_4^2,
\end{equation}
where $\chi$ and $\xi$ are periodic coordinates, and one of the four functions can be fixed by a choice of gauge.

We assume that at least the $\chi$ circle can have antiperiodic boundary conditions for fermions. In this case, the topologically distinct types of solutions are those for which $D$ goes to zero as we decrease $r$ with $B$ and $C$ remaining finite, and those for which $B$ or $C$ goes to zero with $D$ remaining finite. Solutions in which the $\xi$ circle pinches off are related to solutions for which the $\chi$ circle pinches off by $B \leftrightarrow C$.

In order to discuss the details of finding solutions, it will be useful to focus on one particular type of solution, for which the $\chi$ circle of the torus pinches off at some radius $\rho_0 > 0$. We will describe our methods in detail for this case, but the steps for all other cases are completely analogous.

\subsubsection*{Solutions with contractible $\chi$: equations}

To begin, we choose a gauge for the metric ansatz (\ref{genpar}) that is convenient for the present case. To investigate solutions with a degenerating $\chi$ cycle, we write the metric (\ref{genpar}) in the form
\begin{equation}
ds^2= \ell^2 \left\{ \frac{1}{\rho^2}\frac{1}{q(\rho)} d\rho^2   + \rho^2 q(\rho) e^{u(\rho)} d\chi^2 + \rho^2 e^{v(\rho)} d\xi^2 + \rho^2 d\Omega_d^2 \right\},
\label{horo_metric_rescaled}
\end{equation}
where $\rho$ has been defined to be dimensionless and $\chi$ and $\xi$ have identifications $\chi \sim \chi + 2\pi$ and $\xi \sim \xi + 2\pi $. We are interested in solutions where $q(\rho)$ vanishes at some $\rho = \rho_0 > 0$. To avoid a conical singularity at this point, we require
\be
\label{smooth}
 q'(\rho_0) = {2 \over \rho_0^2} e^{-{1 \over 2} u(\rho_0)} \; .
\ee

From the ansatz (\ref{horo_metric_rescaled}), Einstein's equations (\ref{EE}) give three independent equations for the three undetermined functions in the metric. Taking a linear combination of the $\xi \xi$ and $\theta \theta$ component gives
\begin{equation}
2(d+3)\rho^3 q v'+\rho^4u' q v'+2\rho^4q' v'+2\rho^4 q v''+\rho^4 q v'^2+4(d-1)
= 0.
\label{chitheta}
\end{equation}
Next, a linear combination of the $\chi \chi$, $\theta \theta$ and $\xi \xi$ component gives
\begin{equation}
\rho^4 u' q v'+\rho^4 q' v'+2(d+2)(d-1)-2(d+2)(d+1)\rho^2 q-2(d+1)\rho^3
q'+2(d+2)(d+1)\rho^2 = 0,
\label{chithetaxi}
\end{equation}
Finally, combining the $\rho \rho$ and $\chi \chi$ components, we obtain
\begin{equation}
4 v'-2(d+1) u'-\rho u' v'+2\rho v''+\rho v'^2 = 0.
\label{rhochi}
\end{equation}
These equations, together with the constraint (\ref{smooth}), have a symmetry
\be
\label{scale}
v \to v + v_0
\ee
related to the fact that the geometry is a periodic identification in the $\xi$ direction of a solution with noncompact $\xi$.

\subsubsection*{Solutions with contractible $\chi$: boundary conditions}
\label{innerbc}

We would like to solve the equations above subject to the boundary conditions that the boundary metric describes a particular geometry and that the metric in the interior is smooth. In practice, it is simplest to impose all boundary condition at the point $\rho = \rho_0$ and then read off the boundary metric. Looking at the limit of equations (\ref{chithetaxi}) and (\ref{rhochi}) at $\rho=\rho_0$, we find that
\bea
q'(\rho_0) &=& {(d-1) + (d+2) \rho_0^2 \over \rho_0^3} \cr
v'(\rho_0) &=& {-2(d-1) \over (d-1)\rho_0 + (d+2) \rho_0^3}
\eea
The value of $u(\rho_0)$ is fixed in terms of $q'(\rho_0)$ by the smoothness equation (\ref{smooth}). Finally, $q(\rho_0)=0$ by definition, so the only free parameters are $\rho_0$ and $v(\rho_0)$. Solutions with different $v(\rho_0)$ are all related by the scaling symmetry (\ref{scale}), so numerically we need only find solutions for all possible $\rho_0$ with some fixed $v(\rho_0)$. For all solutions, we find that $q(\infty) = 1$, consistent with the asymptotically locally AdS form (\ref{alads}). The parameters of the boundary metric may be read off for a particular solution by
\bea
{2 \pi R_\chi \over R_{S^d}} &=& 2 \pi e^{u_\infty/2} \cr
{2 \pi R_\xi \over R_{S^d}} &=& 2 \pi e^{v_\infty/2} \; .
\eea
Since $v_\infty$ can be freely varied by the scaling symmetry, the radius of the $\xi$ circle is just a free parameter.

\subsubsection*{Solutions with contractible $\chi$: results}

\begin{figure}[t]
  \begin{center}
 \scalebox{0.4}{ \includegraphics{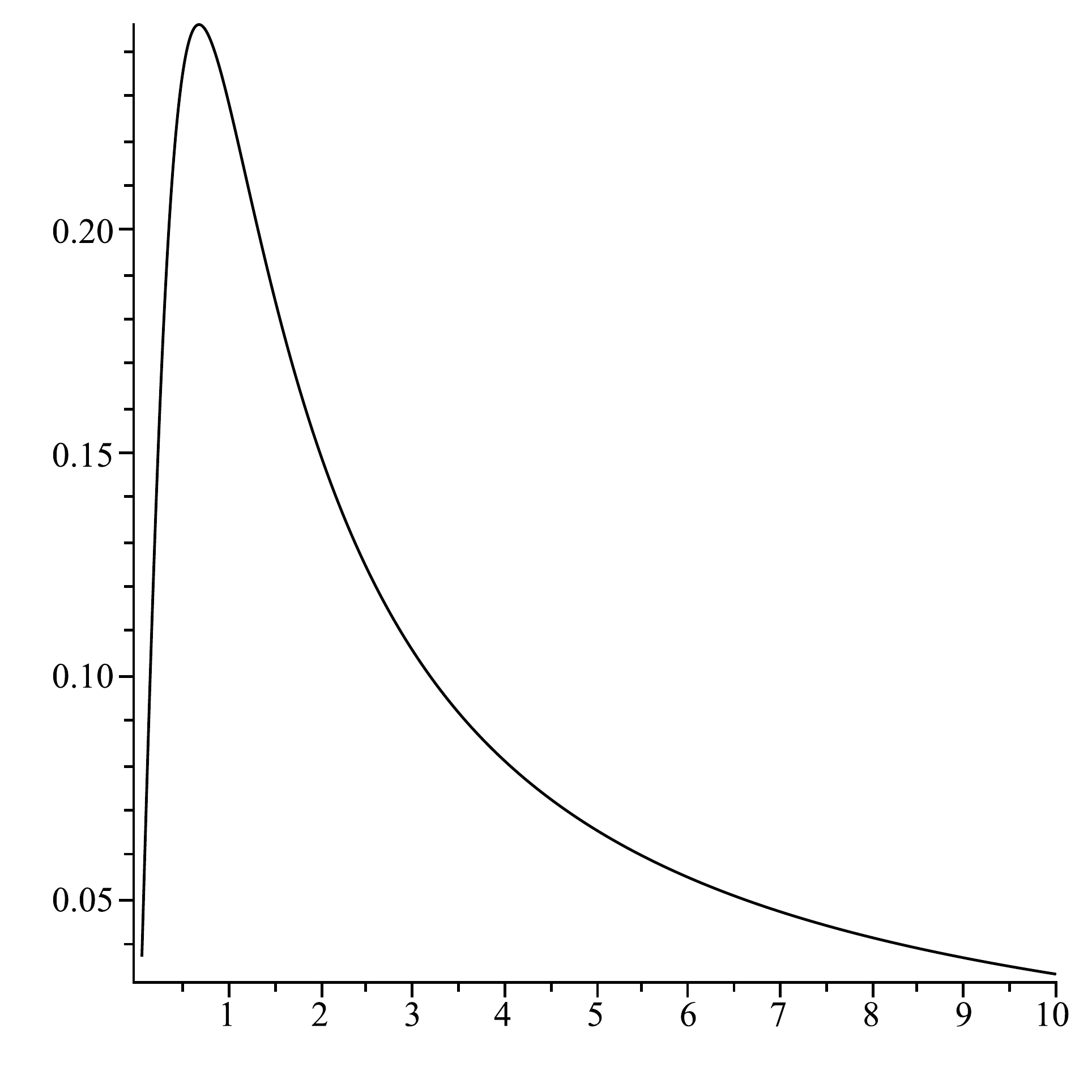}}
  \end{center}
   \caption{A plot of $R_\chi H$ vs $\rho_0$.}
   \label{Rchi}
\end{figure}

The ratio $R_\chi/R_{S^d}$ is plotted against the parameter $\rho_0$ (for the case $d=4$) in Fig. \ref{Rchi}.

The general features of this curve are expected (see e.g. \cite{BalasubramanianLarjoSimon2005}). First, there is a maximum value $(R_\chi)_{max}/R_{S^d} \approx 0.246$ at $\rho_0 \approx 0.672$. For larger $(R_\chi)_{max}/R_{S^d}$, solutions with a contractible $\chi$ circle do not exist.  For $R_\chi < (R_\chi)_{max}$, there are two distinct solutions, distinguished by the value of the radius $\rho=\rho_0$ at which the spacetime smoothly caps off.

We have also obtained numerical results for $d=2$ and $d=3$, and these are completely analogous. In these case, the maximum value of $R_\chi$ for which solutions of this type exist is $(R_\chi)_{max}/R_{S^d} \approx 0.534$ for $d=2$ and $(R_\chi)_{max}/R_{S^d} \approx 0.333$.

\subsubsection*{Solutions with contractible $\xi$: results}

In the case where the boundary conditions for fermions on the $\xi$ cycle are also periodic, we can also have solutions for which the $\xi$ circle is contractible. These are simply related to the solutions just discussed by the exchange $\chi \leftrightarrow \xi$, so they exist whenever $(R_\xi)/R_{S^d} \le (R_\xi)_{max}/R_{S^d}$ where $(R_\xi)_{max}/R_{S^d}$ is $0.246$, $0.333$, or $0.534$ for $d=4,3,2$ respectively.

\subsubsection*{Solutions with contractible $S^d$}

The remaining class of solutions are those for which the $S^d$ contracts to a point at some radius with the $\chi$ and $\xi$ circles having a positive size everywhere. These solutions exist for any values of the $R_\xi$ and $R_\chi$, since they arise by periodic identification from solutions with a boundary geometry $S^d \times R^2$. The solution without identification is dual to the CFT on $S^d \times R^2$, and in this case, we expect that the geometry with least action will not break the symmetries of the $R^2$ or $S^d$. The solution should then take the form
\begin{equation}
\label{global}
ds^2= \ell^2 \left\{ A(\rho) d\rho^2   + B(\rho)( d\chi^2 + d\xi^2 ) + \rho^2 d\Omega_d^2 \right\} \; .
\end{equation}
With this ansatz, Einstein's equations determine $A(\rho)$ in terms of $B(\rho)$ algebraically as
\[
A = { \rho^2 (B')^2 + 4d \rho B B' + 2d(d-1)B^2 \over 2 B^2 ((d+1)(d+2) \rho^2 + d(d-1))}
\]
With this identification, Einstein's equations reduce to a single second order equation for $B$. Each term is cubic in $B$, so we have an scaling symmetry
\[
B \to \alpha B \; .
\]
The solutions we are interested in have positive $B$ for all $\rho \ge 0$, so we can use the scaling symmetry to set $B(0)=1$. The equations of motion then fix $B'(0)=0$, and the rest of the solution is uniquely determined. For large $\rho$, we find that $B(\rho)$ behaves as $c_d \rho^2$, where $c_d$ is a dimension-dependent constant.

To obtain solutions corresponding to the field theory on $S^d \times T^2$ with torus periodicities $2 \pi R_\chi$ and $2 \pi R_\xi$ and sphere size $R_{S^d}$, we identify
\[
\chi \to \chi + {2 \pi R_\xi \over c_d R_{S^d}} \qquad \qquad \chi \to \chi + {2 \pi R_\xi \over c_d R_{S^d}} \; .
\]

\subsubsection*{Phase diagram}

\begin{figure}[t]
 \begin{center}
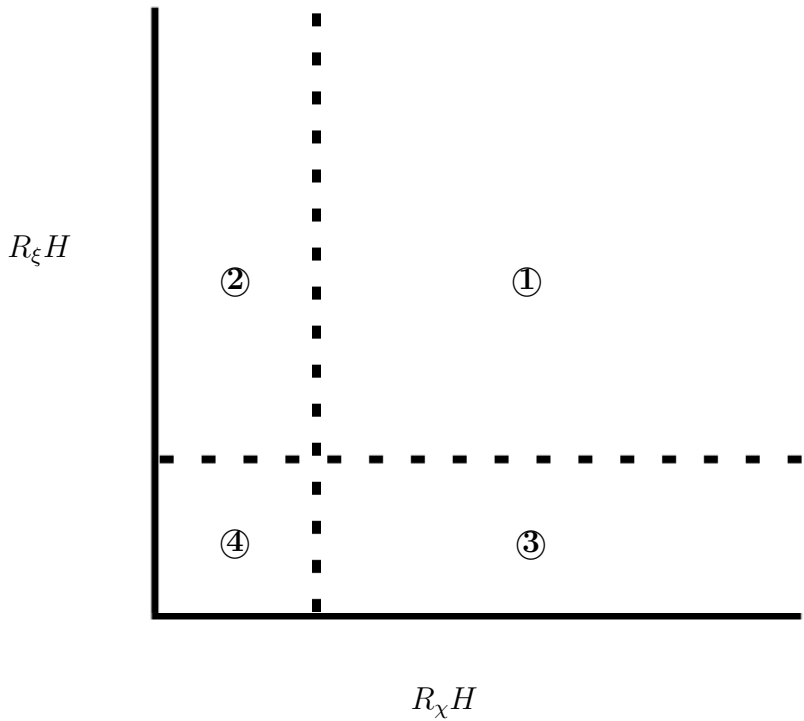
\end{center}
\caption{A plot showing qualitatively distinct regions of the phase diagram.  The dotted lines represent the maximum radius (see Fig. \ref{Rchi}) where the bubble solution can exist.}
\label{phase_reg}
\end{figure}

Taking into account the three types of solutions, the phase diagram can be divided into 4 regions (see Fig.~\ref{phase_reg}) depending on the number and type of solutions.  The large radii region 1 has only one possible solution, whereas small radii region 4 will have a total of 5 solutions to compare (the contractible $S^5$ solution, two solutions where the $\chi$ direction pinches and two solutions where the $\xi$ direction pinches).  The intermediate regions 2 and 3 each have 3 possible solutions to choose from. In the regions with more than one possible solution, we need to compare the various possibilities to determine the one with lowest action. This is dual to the field theory state with lowest free energy.

Comparing the actions is simplified significantly by the fact that the non-contractible circles (one or two in each geometry) arise by periodic identification of solutions in which these directions are non-compact. For any such periodic direction, the action is simply proportional to the radius of the circle. Thus, when comparing solutions for which the $\chi$ circle is contractible to solutions in which the sphere contracts, the ratio of the two actions will be independent of the radius of the $\xi$ circle, which is non-contractible for both cases. So any phase boundary between these two types of solutions will be parallel to the $R_\xi$ axis. Similarly, any phase boundary between contractible $\xi$ solutions and contractible $S^d$ solutions will be parallel to the $\chi$ axis.

Numerically, we find that for sufficiently large $R_\xi$, the solutions with contractible $\chi$ have lower action than the contractible $S^d$ solutions for
\be
R_\chi /R_{S^d} <  \left\{ \ba{ll} 0.239 & \qquad d=4 \cr 0.317 & \qquad d=3 \cr 0.474 &\qquad d=2 \ea \right.
\ee
These phase boundaries are just below the radii
\be
R_\chi /R_{S^d} < \left\{ \ba{ll} 0.246 & \qquad d=4 \cr  0.333 & \qquad d=3 \cr 0.534 &\qquad d=2 \ea \right.
\ee
at which the contractible $\chi$ solutions begin to exist. In the case where the $\xi$ circle is also antiperiodic, we have (by symmetry) a phase boundary between contractible $\xi$ solutions and contractible $S^4$ solutions at $R_\xi/R_{S^d} \approx 0.239$ ($d=4$), $R_\xi/R_{S^d} \approx 0.317$ ($d=3$) or $R_\xi/R_{S^d} \approx 0.474$ ($d=2$), for sufficiently large $R_\chi$.

In region 4, we have five distinct solutions with the same boundary conditions.  We find that the solutions for which either the $\chi$ or $\xi$ directions pinch off at the \emph{smaller} value of the radial coordinate $\rho=\rho_0$ never have the lowest action. For solutions in which the $\xi$ or $\chi$ circle is contractible, the one with lower action is the one for which the corresponding boundary circle has smaller radius. This results in an additional phase boundary along $R_\chi = R_\xi$ which begins at the origin and extends to the intersection of the previous phase boundaries.

\begin{figure}[t]
 \begin{center}
 \def \svgwidth{3.2 in}
 \subfloat[Both cycles anti-periodic.]{\label{anti}  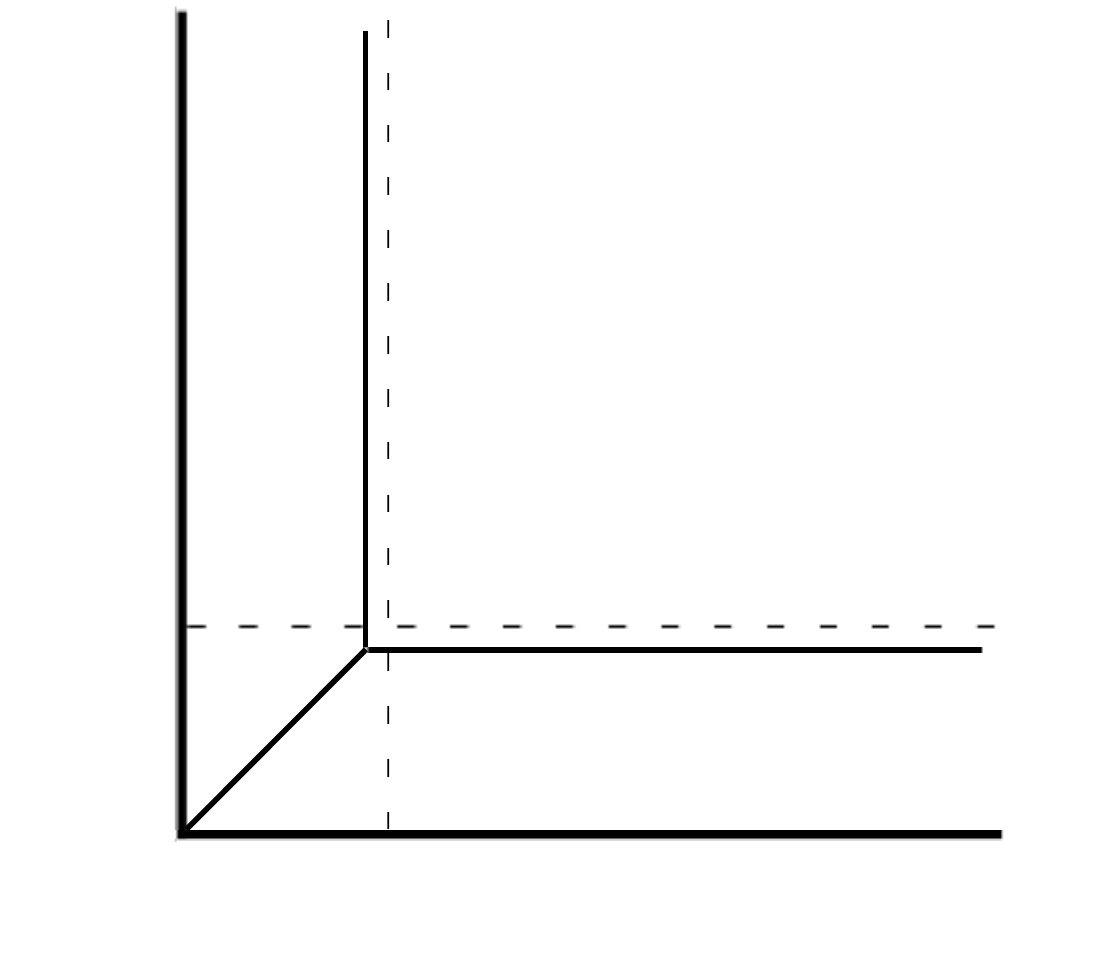 }
 \def\svgwidth{3 in}\subfloat[$\xi$ cycle periodic.  The tick mark is at $R_\chi H = 1/d$.]{\label{per}  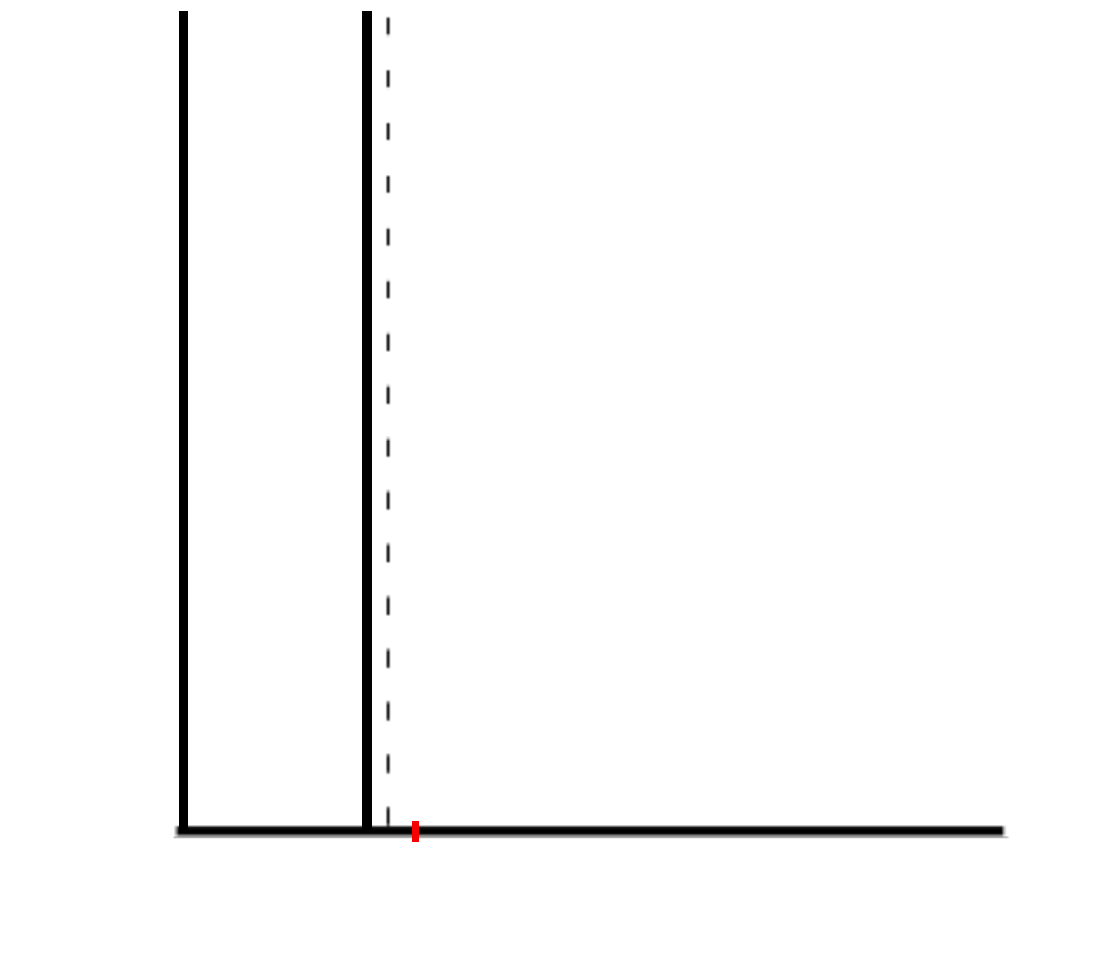 }
  \end{center}
 \caption{Phase diagram for the theory defined by $CFT_{d+2}$ on $T^2$.  Dotted lines represent the maximum radii where solutions with contractible $S^1$ exist.}
 \label{phase_diagram}
 \end{figure}

The phase diagram we find is shown in Fig. \ref{anti}.

\subsubsection*{Critical temperature}

We now interpret our results in terms of confining gauge theory on de Sitter spacetime. For $d=4$, the results above indicate that the $T^2$ compactified CFT undergoes a deconfinement transition at critical de Sitter temperature
\[
T_{dS} = {H \over 2 \pi} = {1 \over 2 \pi R_{S^5}} = {0.239 \over 2 \pi R_\chi} = 0.239 \; T_c  \qquad \qquad (d=4)\; ,
\]
where $T_c$ is the Minkowski space transition temperature. This is close but not equal to the critical de Sitter temperature of $1/4$ for a confining theory defined by compactifying a CFT on $S^1$. In $d=2$ and $d=3$, the results are that
\beas
T_{dS} &=& 0.474 \; T_c  \qquad \qquad (d=2) \cr
T_{dS} &=& 0.317 \; T_c  \qquad \qquad (d=3) \cr
\eeas
which again are close to the the result $T_{dS} = T_c/d$ for confining gauge theory defined by CFT on $S^1$.

\subsection{CFT on $S^2 \times S^d$}

In this section, we consider the case of confining gauge theory defined by $S^2$ compactification of a CFT. For a $(d+2)$-dimensional CFT compactified on $S^2$, the physics on de Sitter space will be obtained by analytic continuation from the physics of the CFT on $S^2 \times S^d$. We will denote the $S^2$ and $S^d$ radii by  $R_{S^2}$ and $R_{dS}$.  Conformal invariance guarantees that the physical properties of this theory on de Sitter space with acceleration parameter $H = 1/R_{dS}$ depend only on the dimensionless parameter $R_{S^2} H$ or equivalently $R_{S^2} / R_{dS}$. Thus, the ``phase diagram'' will be one-dimensional in this case.

Again, we will assume that the field theory state does not spontaneously break any of the geometrical symmetries. Thus, the field theory state and the corresponding gravity solution should possess an $SO(d+1) \times SO(3)$ symmetry.

The most general dual metric with this symmetry can be put in the form:
\be
ds^2 = \ell^2(\frac{dr^2}{r^2} + f(r)d\Omega_2^2 + g(r)d\Omega_d^2)  \; .
\label{s2s4}
\ee

In this case, the topologically distinct types of solutions are those for which $f$ goes to zero as we decrease $r$ with $g$  remaining finite (these correspond to the confined phase), and those for which $g$ goes to zero with $f$ remaining finite (these correspond to the deconfined phase).

Einstein's equations (\ref{EE}) give us two independent ODEs for the two functions $f$ and $g$,
\beas
 r^2(f')^2g^2 + {1 \over 2} d(d-1) r^2 f^2(g')^2 + 2 d r^2 fg  f' g' \qquad  \qquad && \cr
 - 2(d+1)(d+2) f^2 g^2 -  2 d(d-1) f^2 g - 4fg^2 &=& 0 \label{ODE0} \cr
 2d r^2f^2gg'' + 2 r^2 ff''g^2  +d r^2ff'gg' - r^2(f')^2g^2 + {1 \over 2} d (d-3) r^2f^2(g')^2 \qquad  \qquad && \cr
  +2rff'g^2  + 2d rf^2gg' - 2d(d-1)f^2g -2(d+1)(d+2) f^2g^2 &=& 0 \label{ODE1} \\
\eeas
From these, we can eliminate either $f''$ or $g''$ by taking a linear combination of the second equation and the derivative of the first equation. Thus we will require a total of three boundary conditions.

We will consider solutions for which either $f$ or $g$ vanish as we decrease $r$ to some $r=r_0$. Using the scaling symmetry
\[
\tilde{f}(r) = f(\alpha r) \qquad \qquad \tilde{g}(r) = g(\alpha r)
\]
we are free to set $r_0 = 1$. The equations of motion imply that the function which vanishes at $r=1$ also has vanishing derivative at this point. Furthermore, the second derivative at this point is determined by the equations to be 2, and this ensures that the geometry is smooth at $r=1$.

The value of the nonvanishing function at $r=1$ gives the final boundary condition, after which the solution is completely determined. For each type of solution, we can numerically determine field theory parameter $R_{S^2} / R_{dS}$ as a function of $f(1)$ or $g(1)$ using
\[
R_{S^2} / R_{dS} = \lim_{r \to \infty} \sqrt{f(r) \over g(r)} \; .
\]

\subsubsection*{Results and phase structure}

We have carried out the numerics for the cases $d=2$, $d=3$ and $d=4$. As an example, our results for the relation between $f(1) \equiv f_0$ or $g(1) \equiv g_0$ and $f_\infty/g_\infty = R^2_{S^2} / R^2_{dS}$ are plotted in figure \ref{s2s4} for $d=4$.


 \begin{figure}
  \begin{center}
 \scalebox{0.4}{ \includegraphics{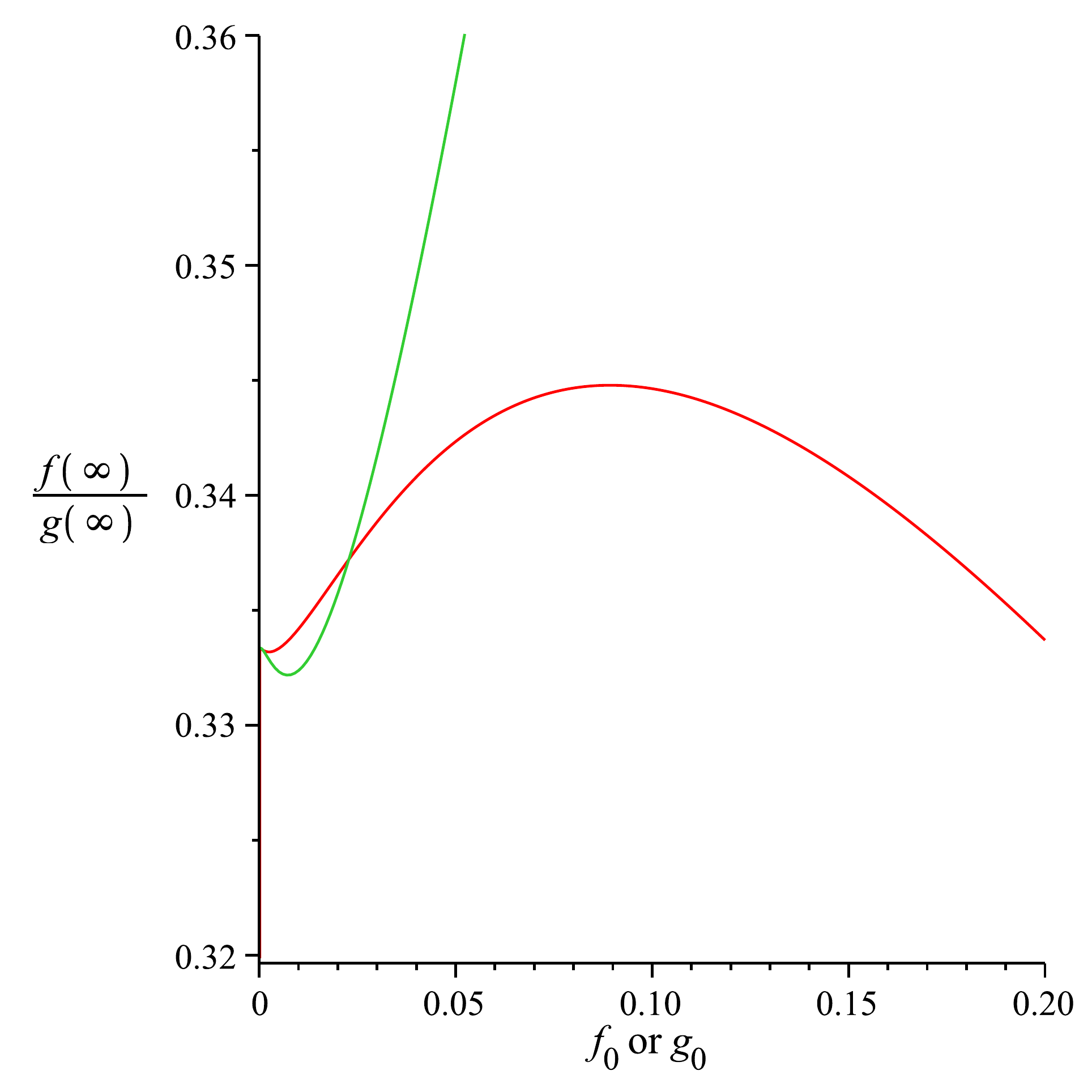}}
  \end{center}
   \caption{Relation between $f_0$ or $g_0$ and $f_\infty/g_\infty = R^2_{S^2} / R^2_{dS}$ for $d=4$.}
   \label{s2s4}
\end{figure}

In each case, there is a range of values for the field theory parameter $R_{S^2} / R_{dS}$ (vertical axis) where multiple solutions exist. Here, we need to compare the action for the various possible solutions to determine which is smallest. For the $d=2$ case, the boundary geometry is $S^2 \times S^2$, and we find that the preferred bulk geometry is always the one for which the smaller of the two spheres is contractible in the bulk. Thus, we have a phase transition at
\[
R_{S^2}/R_{dS} = 1 \qquad \qquad  (d=2)
\]

For $d=3$, we find that solutions for which the $S^2$ is contractible exist in the range $R_{S^2} / R_{dS} \le 0.731$. They have lower action than the contractible $S^3$ solutions below the critical value
\[
R_{S^2}/R_{dS} \approx 0.721 \qquad \qquad  (d=3) \; .
\]

For $d=4$, we find that solutions for which the $S^2$ is contractible exist in the range $R_{S^2} / R_{dS} \le 0.587$. They have lower action than the contractible $S^4$ solutions below the critical value
\[
R_{S^2}/R_{dS} \approx 0.585 \qquad \qquad  (d=4) \; .
\]

\subsubsection*{Critical temperature}

We now interpret our results in terms of confining gauge theory on de Sitter spacetime. The Minkowski space transition temperatures for the $S^2$ compactified CFT are given in (\ref{S2trans}). The de Sitter temperature for which a phase transition will occur is given in terms of the critical $S^d$ radius of the Euclidean theory by $T_{dS} = H/(2 \pi) = 1/(2 \pi R_{S^d})$. Expressing our results from this section in terms of the Minkowski-space deconfinement temperatures from section 3, we find that the transitions occur at
\beas
T_{dS} &=&  0.474 \; T_c \qquad \qquad (d=2) \cr
T_{dS} &=&  0.328 \; T_c \qquad \qquad (d=3) \cr
T_{dS} &=&  0.239 \; T_c \qquad \qquad (d=4)
\eeas
As for the $T^2$ compactified theory, we find that the critical de Sitter temperature is below the Minkowski space deconfinement temperature by a factor which is close but not equal to the dimension $d$ of the de Sitter spacetime.

\section{Discussion}

In this note, we have shown that a variety of confining gauge theories with gravity dual descriptions undergo deconfinement transitions induced by cosmological acceleration at a critical value of the Hubble parameter. Quantitatively, we find that the transition occurs when the de Sitter temperature reaches a critical value close to $T_c/d$, where $T_c$ is the Minkowski-space deconfinement temperature for the same field theory, and $d$ is the dimension of the de Sitter spacetime. The fact that the relation $T_{dS}/T_c \approx 1/d$ holds to within a few percent for all cases suggests that it may hold more generally, perhaps even for realistic confining gauge theories with no known gravity dual.\footnote{On the other hand, as we show in appendix B, if there exist confining gauge theories arising from compactification of $CFT_{d+n}$ on $T^n$ for arbitrary values of $n$, the ratio $T_{dS}/T_c$ must go to zero in the limit $n \to \infty$. However, it is unclear whether consistent conformal field theories exist above six dimensions.}

The existence of such phase transitions may be relevant for our understanding of the inflationary phase of early universe cosmology. Here, either the QCD degrees of freedom or some other non-abelian gauge theory (e.g. a grand unified theory) likely form part of the effective field theory relevant during inflation. Since the Hubble parameter changed from a large value during inflation to a very small value after inflation, it is plausible that during inflation, the non-abelian gauge theory degrees of freedom were in a deconfined phase, and that a confinement transition of the type studied in this paper occurred towards the end of inflation.\footnote{For QCD, this would be a sharp crossover rather than a strict phase transition.} During reheating, when the energy stored in the inflaton field is transferred to other degrees of freedom, we might then have a further transition back to a deconfined phase, followed finally by a third transition to a confined phase as the universe cooled. Alternatively, reheating may have occurred while the non-abelian gauge theory degrees of freedom were still deconfined. In this case, we would have only a single transition from a deconfined to a confined phase after reheating. It would be interesting to understand whether these two possible scenarios result in distinct observable consequences.

\appendix

\section{Minkowski space deconfinement temperature for $CFT_{d+2}$ compactified on $S^2$.}

To determine the Minkowski-space deconfinement temperature for the $S^2$ theory, we need to study asymptotically AdS solutions with boundary geometry $S^2 \times S^1 \times R^{d-1}$. As the $S^1$ radius is decreased relative to the $S^2$ radius, we expect a phase transition between a solution with $S^2$ contractible in the bulk and a solution with contractible $S^1$. For $d=2$, the relevant geometries were analyzed in \cite{Copsey2006}. We review those results presently and extend them to more general $d$.

\subsubsection*{Solutions with contractible $S^2$}

Solutions for which the $S^1$ is not contractible in the bulk exist for any value of the $R_{S^2}$ since they can be obtained by periodic identification from solutions with a boundary geometry $S^d \times R^d$. The solution without identification is dual to the CFT on $S^2 \times R^d$; in this case, we expect that the geometry with least action will not break the symmetries of the $R^d$ or $S^2$. The metric for such a solution may be written as
\begin{equation}
\label{global}
ds^2= \ell^2 \left\{ A(r) dr^2   + B(r)( d w^2 + dx_i dx_i) + r^2 d\Omega_2^2 \right\} \; .
\end{equation}
With this ansatz, Einstein's equations determine $A(\rho)$ in terms of $B(\rho)$ algebraically as
\[
A = {{d(d-1) \over 4} r^2 (B')^2 + 2d r B B' + 2 B^2 \over B^2 (2 +  (d+1)(d+2) \rho^2)}
\]
With this identification, Einstein's equations reduce to a single second order equation for $B$. Each term is cubic in $B$, so we have an scaling symmetry
\[
B \to \Lambda B \; .
\]
The solutions we are interested in have positive $B$ for all $\rho \ge 0$, so we can use the scaling symmetry to set $B(0)=1$. The equations of motion then fix $B'(0)=0$, and the rest of the solution is uniquely determined. For large $\rho$, we find that $B(\rho)$ behaves as $\beta_d \rho^2$, where $\beta_d$ is a dimension-dependent constant.

To obtain solutions corresponding to the field theory on $S^2 \times S^1 \times R^{d-1}$ with $S^1$ periodicity $2 \pi R_{S^1}$ and sphere size $R_{S^2}$, we identify
\be
\label{per}
w \to w + {2 \pi R_{S^1} \over \sqrt{\beta_d} R_{S^2}} \; .
\ee

\subsubsection*{Solutions with contractible $S^1$}

\begin{figure}
  \begin{center}
 \scalebox{0.4}{\includegraphics[angle=-90]{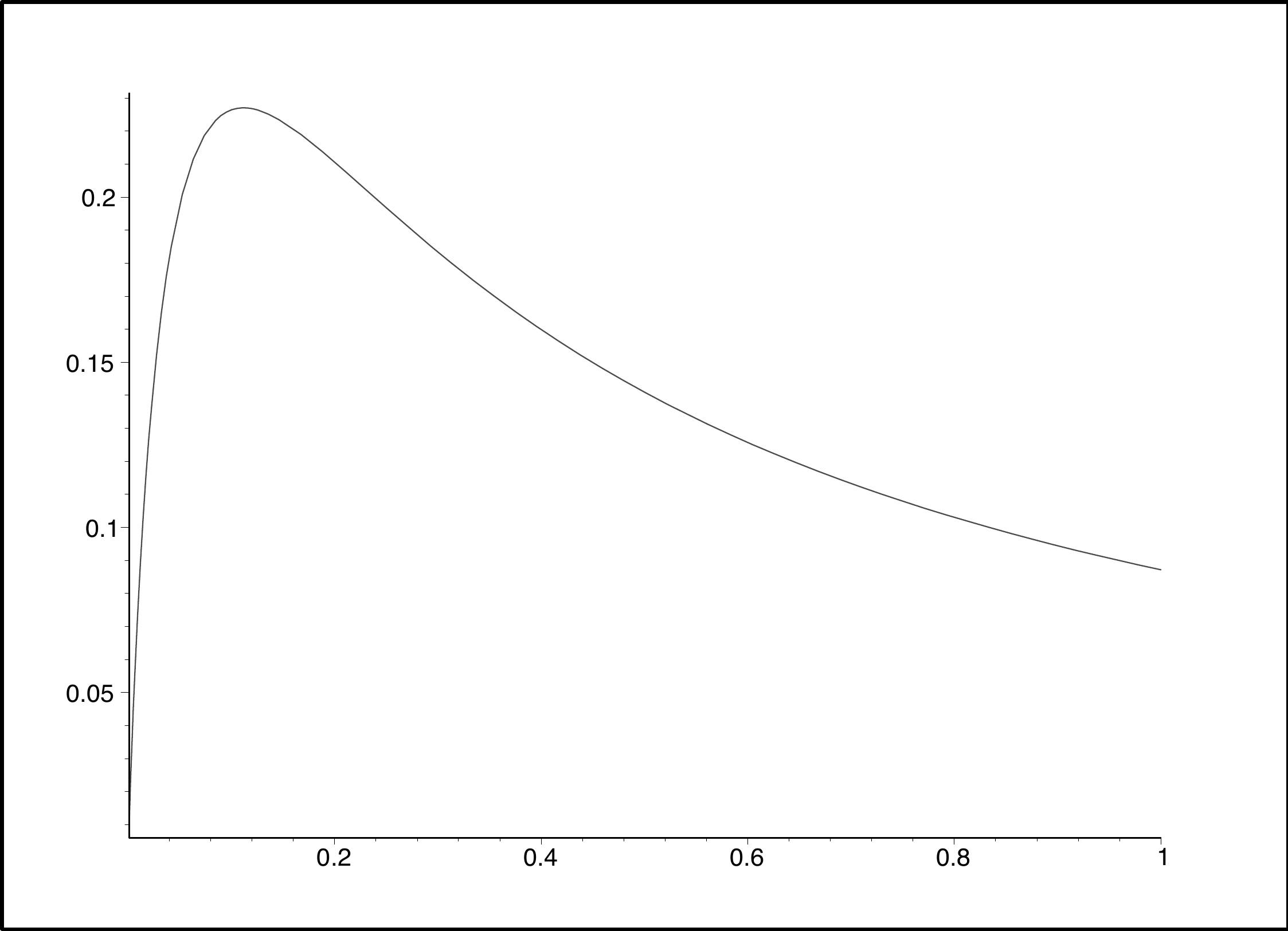}}
  \end{center}
   \caption{Relation between $g_0$ (horizontal axis) and $f_\infty/g_\infty = R^2_{S^1} / R^2_{S^2}$ (vertical axis) for solutions with boundary geometry $S^2 \times S^1 \times R^3$ and contractible $S^1$.}
   \label{s2s1r3}
\end{figure}

Solutions for which the $S^1$ contracts smoothly in the bulk cannot be obtained by periodic identification. For this case, we use the ansatz

\[
ds^2 = \ell^2 ( {dr^2 \over r^2} + f(r) d \theta^2 + h(r) dy_i dy_i + g(r) d \Omega_2^2)
\]
where $\theta$ is defined to be periodic,  $\theta \sim \theta + 2\pi$. Einstein's equations give three independent equations for the three functions $f,g$, and $h$, one of them first order in derivatives, and the other two second order. We are interested in solutions where $f(r)$ vanishes at some $r = r_0 > 0$. The equations have three independent scaling symmetries (related to conformal invariance, and reparametrizations of $y_i$ and $\theta$),
\beas
\tilde{f}(r) &=& A f(\gamma r) \cr
\tilde{g}(r) &=& g(\gamma r) \cr
\tilde{h}(r) &=& B h(\gamma r)
\eeas
We can choose $\gamma$ to set $r_0=1$. To avoid a conical singularity, we then require that
\[
f(r) = (r-1)^2 + {\cal O}((r-1)^3) \;
\]
which we can arrange by demanding $f'(1)=0$ and using the scaling symmetry to fix the coefficient of the quadratic term. The remaining scaling symmetry may be used to fix $h(1)=1$. With these choices, the equations of motion imply that $g'(1)=h'(1)=0$. With all these conditions, there remains only a single free parameter $g(1) \equiv g_0$ in terms of which the solution is completely determined.
The parameters of the boundary metric may be read off for a particular solution by
\be
\label{ratio}
\frac{R_{S^1}}{R_{S^2}} = \sqrt{ \frac{f_\infty}{g_\infty}}
\ee
For $d=4$, our results for the ratio $f_\infty/g_\infty$ as a function of $g_0$ are plotted in figure \ref{s2s1r3}.

\subsection*{Comparing actions}

From the figure, see that a pair of solutions with contractible $S^1$ exist for $R_{S^1} / R_{S^2}$ below some critical value (0.477 in $d=4$, 0.502 in $d=3$ and 0.534 in $d=2$). In this range, we need to compare the action for these two solutions with the action for the solution with contractible $S^2$. We find that the preferred bulk geometry (with minimum action) is the one with contractible $S^2$ for $R_{S^1} / R_{S^2} > 0.436$ in $d=4$, $R_{S^1} / R_{S^2} > 0.453$ in $d=3$, or $R_{S^1} / R_{S^2} > 0.474$ in $d=2$. Below this value the contractible $S^1$ solution with the larger value of $g_0$ provides the minimum action solution. Thus, we have a phase transition at
\beas
R_{S^1} / R_{S^2} &=& 0.474 \qquad \qquad  (d=2) \cr
R_{S^1} / R_{S^2} &=& 0.453 \qquad \qquad  (d=3) \cr
R_{S^1} / R_{S^2} &=& 0.436 \qquad \qquad  (d=4)
\eeas
A detailed discussion of the action comparison may be found in the next appendix.

\section{Evaluating the action}

In this section, we provide some details on evaluating the action via the regularized expression (\ref{action}). We take as an example the case of boundary geometry $S^2 \times S^1 \times R^1$. Here, we have used one ansatz
\be
\label{fghsol}
ds^2 = \ell^2 ( {dr^2 \over r^2} + f(r) d \theta^2 + h(r) dy^2 + g(r) d \Omega_2^2)
\ee
for solutions with contractible $S^1$ and another ansatz
\begin{equation}
\label{global2}
ds^2= \ell^2 \left\{ A(r) dr^2   + B(r)( d w^2 + dx^2) + r^2 d\Omega_2^2 \right\} \; .
\end{equation}
for solutions with contractible $S^2$.

For $D=5$, (setting $l=1$ and absorbing a factor $16 \pi G^{(D)}$ into the action) the relevant gravitational action (eq. \ref{action})), including counterterms (eq. \ref{ctaction}) is
\be
\label{action3}
S =  -  \int_{\cal M} d^5 x \sqrt{g} \left\{ {\cal R} + 12 \right\} + \int_{\partial \cal M} d^4 x \sqrt{\gamma} \left\{-2 K + 6 + {1 \over 2} \hat{{\cal R}} \right\} \;,
\ee
Using the fact that the trace part of Einstein's equations (\ref{EE}) imply
\[
{\cal R} = -20 \; ,
\]
the first term simplifies to
\[
S_{bulk} =  \int_{\cal M} d^5 x \{ 8 \sqrt{g} \}
\]
For large $r$, the various components of the metric behave as $g_{rr} \sim 1/r^2$ and $g_{ij} \sim r^2$ for the remaining components, so the integrand in the bulk action grows like $r^3$ (or $r^{D-2}$ in general). This growth should be canceled by the boundary terms, but recovering the finite result from the difference of two very large contributions is challenging numerically. Thus, it is helpful to transfer some of the bulk contribution to the boundary action by adding and subtracting a total derivative. The large $r$ behavior of the metric components implies that the expression $\partial_r (2 \sqrt{\gamma})$ has the same asymptotic behavior as $8 \sqrt{g}$. Thus, the expression
\be
\label{action4}
S =  \int_{\cal M} d^5 x \left\{ 8 \sqrt{g} - 2 \partial_r{ \sqrt{\gamma}} \right\} + \int_{\partial \cal M} d^4 x \sqrt{\gamma} \left\{-2 K + 8 + {1 \over 2} \hat{{\cal R}} \right\} \;,
\ee
is equivalent to the previous expression (\ref{action}), but now the bulk term grows more slowly with $r$.\footnote{We could add and subtract further total derivative terms to cancel off more terms in the asymptotic expansion of the bulk action, but the expression here usually suffices for the numerical evaluation.} For solutions of the form (\ref{global2}), this gives
\be
\label{ABact}
S = \int d\Omega^2 dx dw  \left[ \int_0^{r_M} \left\{8 r^2 \sqrt{A} B -4 r B -2 r^2 \partial_r B \right\} +  r_M^2 B \left\{-{4 \over r_M \sqrt{A}}- 2 { \partial_r B  \over B \sqrt{A}} +8+{1 \over r_M^2} \right\} \right]
\ee
Evaluating this numerically for a solution, we find that the expression still diverges logarithmically as $r_M \to \infty$. Using the asymptotic form for the solution (here $B_2$ is a constant that depends on the solution)
\beas
B(r) &=& \beta \left(r^2+{1 \over 2}+{1 \over r^2}(B_2+{ 1 \over 12} \ln(r))  + {\cal O}(r^{-4}) \right) \cr
A(r) &=& {1 \over r^2}-{2 \over 3 r^4}+{1 \over r^6}({4 \over 9} -2 B_2 - {1 \over 6} \ln(r))+ {\cal O}(r^{-8})
\eeas
we can check analytically that the expression (\ref{ABact}) behaves for large $r_M$ as
\[
S \sim -{1 \over 6} \beta \ln(r_M) + {\rm const} \; .
\]
This divergence may be canceled by adding an extra counterterm
\be
\label{Sct}
S_{ct}^{log} = \int_{\partial \cal M} d^4 x \sqrt{\gamma} \left\{ - {1 \over 48} {\hat{\cal R}}^2 \ln(\hat{\cal R}) \right\} = \int d\Omega^2 dx dw \; r_M^2 B \left\{ -{1 \over 12 r_M^4} \ln \left( {2 \over r_M^2} \right) \right\} \; .
\ee
This counterterm, expressed in terms of the intrinsic geometry of the boundary surface, can be used for any other solution with the same asymptotic behavior, and the resulting regulated expressions can be directly compared.

Our final expression for the action for solutions with contractible $S^2$ is
\[
S_1 = \int d\Omega^2 dx dw s_1
\]
where
\beas
s_1 &=& \int_0^{r_M} \left\{8 r^2 \sqrt{A} B -4 r B -2 r^2 \partial_r B \right\}  \cr
&& \qquad \qquad +  r_M^2 B \left\{-{4 \over r_M \sqrt{A}}- 2 { \partial_r B  \over B \sqrt{A}} +8+{1 \over r_M^2} -{1 \over 12 r_M^4} \ln \left( {2 \over r_M^2} \right) \right\}
\eeas

For solutions of the form (\ref{fghsol}) the expression for the action (\ref{action4}) together with the counterterm (\ref{Sct}) is
\[
S_2 = \int d\Omega^2 dy d \theta s_2
\]
where
\beas
\label{fghact}
s_2 &=&   \int_1^{r_M} \left\{{8 \over r} g f^{1 \over 2} h^{1 \over 2}-2 \partial_r g f^{1 \over 2} h^{1 \over 2} -g f^{-{1 \over 2}} h^{1 \over 2} \partial_r f -g f^{1 \over 2} h^{-{1 \over 2}} \partial_r h \right\} \cr
 && \qquad \qquad  +  g f^{1 \over 2} h^{1 \over 2} \left\{-r_M {\partial_r f \over f} - r_M {\partial_r h \over h}-2r_M {\partial_r g \over g}+8+{1 \over g}-{1 \over 12 g^2} \ln \left({2 \over g}\right) \right\}
\eeas

In the expressions above, both $s_1$ and $s_2$ give finite results in the limit $r_M \to \infty$. However, we should not compare these directly, because the coordinate $w$ has a different periodicity from $\theta$ and the coordinates $y$ and $x$ are not equivalent. From (\ref{ratio}) and (\ref{per}), we see that for a given solution of the form (\ref{fghsol}), contractible $S^2$ solutions with the same $\frac{R_{S^1}}{R_{S^2}}$ are obtained by choosing the periodicity of $w$ to be $2 \pi \sqrt{f_\infty/g_\infty} \sqrt{(r^2/B)_\infty}$. To ensure that we integrate over an equivalent volume in the non-compact field theory directions, we can choose the ranges of $x$ and $y$ so that the proper distance that we integrate along the noncompact field theory direction matches the $S^2$ radius at the same value of $r$. Thus, we choose the range of $y$ to be $\sqrt{g_\infty/h_\infty}$ and the range of $x$ to be $\sqrt{(r^2/B(r))_\infty}$. In the end, to determine which of the solutions has lower action per field theory volume, we must therefore compare
\[
s_1  \sqrt{f_\infty \over g_\infty} \left( {r^2 \over B(r)} \right)_\infty
\]
with
\[
s_2 \sqrt{g_\infty \over h_\infty} \; .
\]

\section{Conformal theories from compactification of $CFT_{4+n}$ on $T^n$.}

In the analysis of this paper, we found that the critical de Sitter temperature (relative to the corresponding Minkowski deconfinement temperature) for a confining theory defined by $CFT_{d+2}$ compactified on $T^2$ is slightly lower than that for a $CFT_{d+1}$ compactified on $S^1$. In this appendix, we consider the generalization to confining theories defined by compactification of $CFT_{d+n}$ on $T^n$, to see whether the critical temperature decreases with $n$.

To investigate this, we focus on the case $d=2$, and calculate the maximum value of $R_{S^1}/R{S^2}$ for which a solution with contractible $S^1$ exists (where the $S^1$ represents one of the periodic directions on the torus). This provides an upper bound for the critical value of $H_{dS} R_{S^1}$.

\begin{figure}[t]
  \begin{center}
 \scalebox{0.5}{ \includegraphics{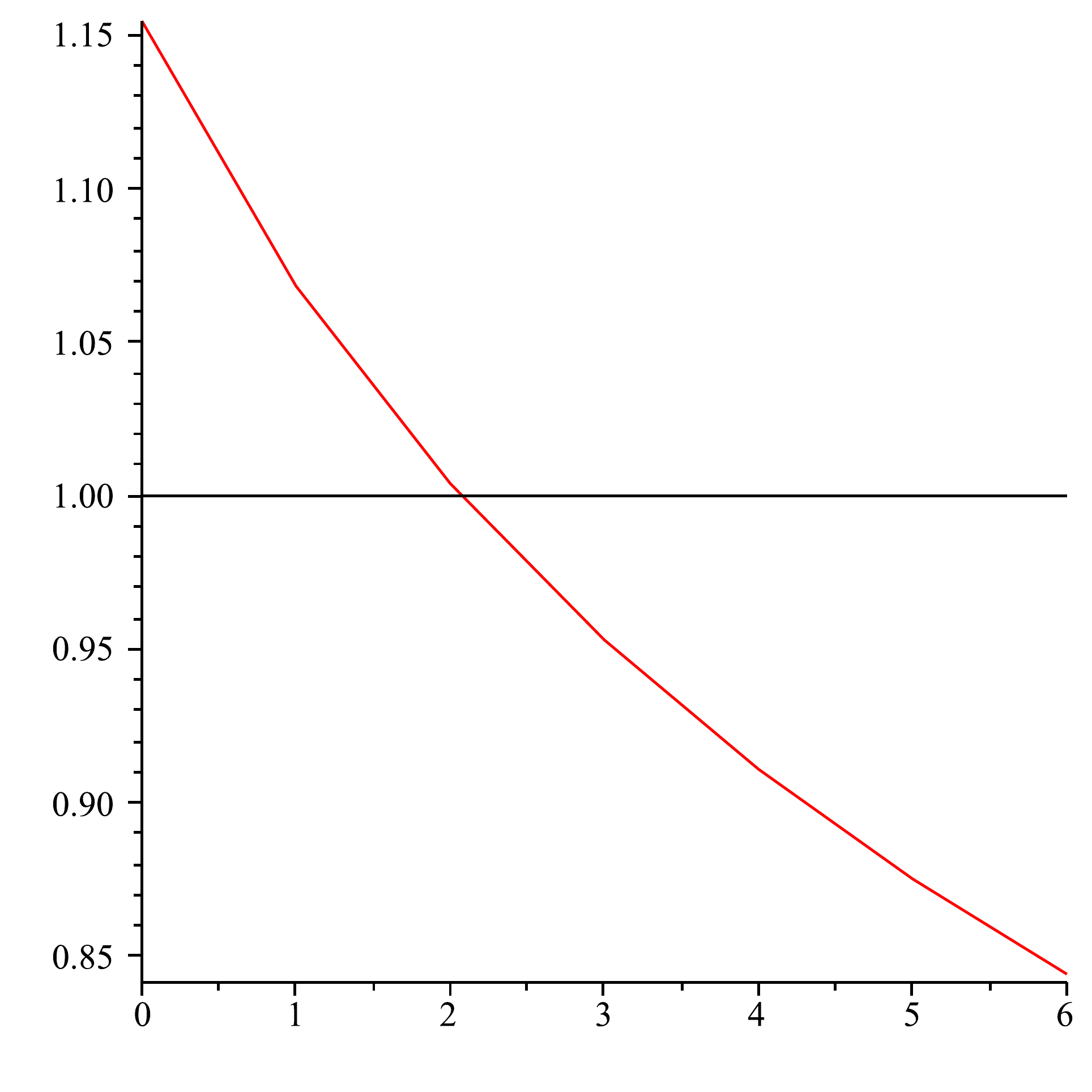}}
  \end{center}
   \caption{A plot of $(R_\chi H)_{max}$ vs $n$, normalized to $1/d$.  The straight line represents where the phase transition would occur if the relation $T_{dS}=T_{Minkowski}/d$ were preserved for arbitrary $n$.}
   \label{Rmax}
\end{figure}

Using all the previously discussed symmetries, we can put a metric with conformal boundary $dS_2\times T^{n+1}$ in the form:

\begin{equation}
ds_{(n)}^2= \frac{\ell^2}{\rho^2}\frac{1}{q(\rho)} d\rho^2   + \frac{\rho^2}{\ell^2} q(\rho) e^{u(\rho)} d\chi^2 +\frac{\rho^2}{\ell^2} e^{v(\rho)} \sum_{i=1}^n d\xi_i^2 + \frac{\rho^2}{\ell^2} d\Omega_2^2,
\end{equation}
with $\displaystyle \xi_i \sim \xi_i + 2\pi R_{\xi_i} H$.
For $n=0$ the solution is exact and we have:
\begin{equation}
R_\chi=\frac{2\rho_0}{1+3\rho_0},
\end{equation}
so that $(R_\chi)_{max}=1/\sqrt{3}$.  For other $n$ the boundary conditions are found to be
\begin{equation}
\begin{cases}
q'(\rho_0) = \frac{1+(n+3)\rho_0}{\rho_0^3}, \\
v'(\rho_0)= -\frac{2}{\rho_0(1+(n+3)\rho_0^2)}.
\end{cases}
\end{equation}
Using these we can integrate the equations of motion to obtain the maximal allowed $H_{dS} R_{S^1}$.  This is shown in Fig.~\ref{Rmax}.  It is apparent that the maximum value is a decreasing function of $n$. On the other hand, the Minkowski space deconfinement temperature is $T_c (2 \pi R_{S^1}) = 1$, independent of the dimension of the torus.

This suggests that ratio of the critical de Sitter temperature to the Minkowski space transition temperature decreases as we increase the dimension of the underlying CFT, keeping the dimension of the compactified theory fixed. We should note, however, that conformal field theories with gravity duals may not exist above six dimensions.

\bibliographystyle{utphys}
\bibliography{dSbibnew}

\end{document}